\documentclass[useAMS,manuscript]{emulateapj}
\usepackage{epsfig}
\usepackage{amssymb}
\usepackage{natbib}
\usepackage{graphicx}

\def\beqn{\vspace{2mm} \begin{eqnarray}}
\def\eeqn{\vspace{2mm} \end{eqnarray}}
\newcounter{parentequation}

\def\m{{\rm\thinspace m}}

\slugcomment{Not to appear in Nonlearned J., 45.}

\shorttitle{Mergers of disks}
\shortauthors{Naab et al.}

\begin{document}


\title{Are disk galaxies the progenitors of giant ellipticals?}

\author{Thorsten Naab$^{1}$ \& Jeremiah P. Ostriker$^{2,3}$} 
\affil{$^1$ Universit\"ats-Sternwarte M\"unchen, Scheinerstr.\ 1, D-81679 M\"unchen, Germany; \texttt{naab@usm.lmu.de}\\
$^2$ Institute of Astronomy, Madingley Road, Cambridge CB3 0HA, UK \\
$^3$ Department of Astrophysics, Peyton Hall, Princeton, USA}



\begin{abstract}
A popular formation scenario for giant elliptical galaxies proposes
that they might have formed from binary mergers of disk
galaxies. Difficulties with the scenario that emerged from earlier
studies included providing the necessary stellar mass and metallicity,
maintaining the tight color-magnitude relation and avoiding phase
space limits. In this paper we revisit the issue and put
constraints on the binary disc merger scenario based on the stellar
populations of disc galaxies. We draw the following conclusions: Low
redshift collisionless or gaseous mergers of present day Milky Way
like disc galaxies do not form present day elliptical
galaxies. Binary mergers of the progenitors of present day Milky Way like 
disc galaxies can have evolved into intermediate mass elliptical galaxies
($M < M_*$) if they have merged earlier than $\approx$ 3-4 Gyrs
ago. Assuming that most present day disk galaxies formed in a similar way to 
the Milky Way model presented here, more massive giant ellipticals in 
general can not have formed from binary mergers of the progenitors of present day disc
galaxies. A major reason for these conclusions is that the mass in metals of
typical disk galaxy is approximately a factor of 4-8 smaller than the
mass in metals of a typical early-type galaxy and this ratio grows to
larger values with increasing redshift. 
\end{abstract}





\section{Introduction}

How did elliptical galaxies form? After more than a decade of high
resolution observations from space and new powerful numerical
simulations, additional insight has been gained but also new and old questions have
been raised. Red bulge-dominated galaxies contain at least half of the stellar mass 
of the universe \citep{1998ApJ...503..518F,2002AJ....124..646H} and adding bulges of 
spiral galaxies the fraction might be as high as 3/4
\citep{2003ApJS..149..289B}. Therefore a consistent theory for
elliptical galaxy formation is of fundamental importance but it is still
unavailable. Independent of the cosmological context, the luminous parts
of massive galaxies must have formed by collecting baryonic material
under the influence of gravity. So, in the most general sense
`mergers' of some type are of course necessarily the precursors of
present day elliptical  systems. The questions are when and how 
they have formed. There are a variety  of possibilities concerning how
this could have happened. Depending on whether the infalling material
has already collapsed and formed stars the galaxy could either accrete
stars or accrete gas which is then turned into stars thereafter within the
galaxy.  In addition, the distribution of sizes and gas fractions of
the accreted material might vary with time. The choice of these
parameters is not completely arbitrary and a detailed theory of
elliptical galaxy formation must be constrained by comparison with the
properties of observed elliptical galaxies and of the progenitor
components.   

It has been known for decades that giant elliptical galaxies are a
homogeneous family of galaxies with old stellar populations
(e.g.
\citealp{1976ApJ...204..668F,1989ARA&A..27..235K,2005ApJ...621..673T,2006ApJ...647L..99D}).
Therefore it 
has been suggested that they all have formed {\it in situ} at early
times \citep{2000ApJ...536L..77B}. Without, at this earlier time, detailed  
knowledge of the, currently popular, hierarchal cosmological models,
the 'monolithic collapse' scenario was designed to explain the early
formation of ellipticals motivated by the \citet{1962ApJ...136..748E}
proposal that the old spheroidal component of the Milky Way formed
during a short period of quasi-radial collapse of gas some $10^{10}$ Gyrs
ago and, in most of their properties, the spheroidal bulges of spiral
and S0 galaxies are indistinguishable from ellipticals of the same
luminosities.   

In that sense, an elliptical/spheroidal galaxy would have 
formed very early, as soon as a finite sufficiently over-dense
region of gas and dark matter decoupled from the expansion of the
universe and collapsed. If, during the proto-galactic collapse phase,
star formation was very efficient, a coeval spheroidal stellar system
could have formed before the gas could have dissipated its kinetic 
and potential energy and would have settled into the equatorial plane, thus avoidng the 
 formation of a disk galaxy \citep{1967ApJ...147..868P, 1969MNRAS.145..405L, 1974MNRAS.169..229L,1973ApJ...179..427S}. Depending 
on the amount of 'turbulent viscosity' and the angular momentum of the
infalling gas, detailed properties like isophotal shapes, rotation, age
and metallicity gradients and even the formation of disk-like
substructures in elliptical galaxies have been computed decades ago
\citep{1975MNRAS.173..671L}.  The 'turbulent viscosity' was 
created by large inhomogeneities and random motions, eventually caused
by independently moving gas clouds. Recent high resolution numerical simulations,  
using modern cosmological models, indicate that indeed there was a
phase of evolution of this type \citep{2007ApJ...658..710N} during
which a small ($\propto 1-2 kpc$), concentrated and massive system was
formed rapidly at high redshifts ($2 \leq z \leq 4$).

At about the same time
\citet{1972ApJ...178..623T,1974IAUS...58..347T,1977egsp.conf..401T} 
investigated the tidal interactions during encounters of spiral
galaxies and proposed that disk galaxies might eventually merge and
form elliptical galaxies. This proposal became particularly attractive
with the advent of hierarchical cosmological theories
\citep{1978MNRAS.183..341W} in which mergers play the dominant role
during the formation and/or evolution of every dark matter halo and
almost every galaxy at some time in its life. In this 'merger scenario'
massive elliptical galaxies could have formed from the gravitational
interaction of already existing massive galaxies that were not of
early type.

\section{The disk merger hypothesis} 

\subsection{Collisionless disc mergers} 
In the early 80's the first fully self consistent  
simulations of mergers of stellar disk galaxies were performed 
\citep{1981MNRAS.197..179G,1983MNRAS.205.1009N}.
Using effective tree algorithms \citep{1986Natur.324..446B,1987ApJS...64..715H,
1989ApJS...70..389B} and more powerful general purpose and special
purpose computers a whole industry of simulations of merging
disk galaxies was created that was designed to answer the 
question of whether they form systems that resemble present day
elliptical galaxies \citep{1988ApJ...331..699B,1990Natur.344..379B,1992ARA&A..30..705B,
1992ApJ...393..484B,1992ApJ...400..460H,2003ApJ...597..893N,2005MNRAS.360.1185J,
2005MNRAS.357..753G,2006MNRAS.369..625N}. 
The global properties of the remnants are in several respects consistent
with observations of giant elliptical galaxies, e.g. equal mass
remnants are triaxial, slowly rotating, anisotropic, have boxy or discy
isophotes \citep{1994ApJ...427..165H,2003ApJ...597..893N}. In
addition, mergers of discs can result in the formation of kinematic
subsystems like kinematically decoupled cores at the centers of
ellipticals  \citep{1991Natur.354..210H,2006astro.ph..6144J} as well
as observed faint structures like shells, loops and ripples at large
radii \citep{1992ApJ...399L.117H}. Unequal mass mergers are more
supported by rotation \citep{1998giis.conf..275B} and have discy isophotes
\citep{2003ApJ...597..893N}. With respect to their intrinsic
structure, all collisionless merger remnants are dominated by box
orbits at their centers and tube orbits in the outer parts
\citep{1998giis.conf..275B}. The total fraction of tube orbits
increases  with the mass ratio of the mergers
\citep{2005MNRAS.360.1185J}. The fraction of disc and box orbits
correlates with the shape and kinematics of the systems and the mix of
discy and boxy isophotal shapes for equal-mass remnants can be
understood by the projected properties of tube orbits in triaxial
potentials \citep{2005MNRAS.360.1185J,2006MNRAS.372..839N}.
\citet{2003ApJ...597..893N} and have argued, based on statistics of
kinematic and photometric properties of equal- and unequal-mass
mergers, that disc mergers (with bulges) can result in intermediate
mass elliptical galaxies. But the objects so formed are not in
agreement with the most massive, boxy and slowly rotating ellipticals.
\citet{2006ApJ...636L..81N} indicate major binary early type mergers could 
be responsible for the slow rotation of the most massive ellipticals 
(see also \citealp{2005MNRAS.359.1379K,2005MNRAS.361.1043G})

Collisionless merger remnants in general have phase space densities and 
surface density profiles that resemble observed ellipticals only if bulges are 
added to the progenitor discs (see
e.g. \citealp{1986ApJ...310..593C}). The phase space densities of pure 
stellar disc mergers are too low \citep{1993ApJ...416..415H,2006MNRAS.369..625N}.
In addition, intrinsic and isophotal shapes
\citep{2005MNRAS.360.1185J,2006ApJ...650..791C} as well as surface
density profiles (\citealp{2005MNRAS.357..753G,2006MNRAS.369..625N}, see  however
\citealp{2006MNRAS.373..632A}) of pure disc mergers are not in
agreement with elliptical galaxies. At this point, however, there
remains the question of how bulges have formed in the first place.       

The agreement of the kinematics of collisionless disc mergers and
observed elliptical galaxies is only good to first order. At higher
order there are disagreements with observed ellipticals. 
The line-of-sight velocity distributions (LOSVD) within 
the effective radius of merger remnants in general show small asymmetric 
deviations from Gaussian shape. They tend to have a steep trailing wing
(\citealp{2000MNRAS.316..315B,2001ApJ...555L..91N,2006MNRAS.372..839N}, see however 
\citealp{2006MNRAS.372L..78G} for the influence of massive bulges in the progenitor disks), 
whereas most observed rotating ellipticals show a steep leading wing in their
LOSVDs \citep{1994MNRAS.269..785B}. Theoretically, axisymmetric,
rotating one-component systems show such a behaviour
\citep{1994MNRAS.268.1019D}. This would indicate that ellipticals are
very simple one component systems that did not form by mergers, which
is, however, unlikely (see e.g. \citealp{2004MNRAS.352..721E}). 
An alternative explanation, based on photometric and kinematical 
observations, is that rotating ellipticals contain embedded 
large scale stellar discs (e.g. \citealp{1988A&A...193L...7B,1989A&A...217...35B,1990ApJ...362...52R,1998A&AS..131..265S,
1999ApJ...513L..25R}). A superposition of two distinct components, 
e.g. a hot spheroidal bulge and a rotationally supported cold disc, 
can also result in a steep leading wing of the LOSVD
\citep{1994MNRAS.269..785B,2001ApJ...555L..91N}.

\subsection{Disc mergers with gas}

Evidently, disc galaxies do not only consist of stars but also an 
interstellar medium (ISM) in the form of gas. In the local universe
for evolved disc galaxies typical gas fractions are 10 - 30 per cent
of the total stellar mass (e.g. \citealp{1997ApJ...481..689M}). Within
the hierachical  paradigm the gas fraction is an increasing 
function with redshift. (see e.g. \citealp{2006MNRAS.370..902K})
Even if the overall gas fractions were relatively small, due to its
dissipative nature gas can change the structure of merger 
remnants significantly (see e.g. \citealp{2006ApJ...648L..21K,2006astro.ph.11328C})
It has been shown by
\citet{1996ApJ...471..115B} that gas accumulating at the center of
merger remnants creates a steep cusp in the central potential well
resulting in a more axisymmetric central shape of the remnants
\citep{1998giis.conf..275B}. At the same time the fraction of stars on
box orbits is significantly reduced and tubes become the dominant
orbit family \citep{1996ApJ...471..115B}. The most reasonable
explanation for this behaviour is that systems with steep cusps in
their potential can not sustain a large population of box orbits
\citep{1985MNRAS.216..467G,1993ApJ...409..563S,1996ApJ...460..136M,1998ApJ...506..686V,1998giis.conf..275B,2006MNRAS.372..839N}.
One interesting consequence of the change of orbit populations is that
the asymmetry of the LOSVD of the stars changes in a way that is
consistent with observations  making it a good dynamical tracer for
the presence of gas during the merger \citep{2006MNRAS.372..839N}. 

Dissipational merging, including star formation, can also overcome
stellar phase space constraints and therefore the a priori inclusion
of a major sheroidal bulge component is not required, and it has been shown by
\citep{2006ApJ...641...21R} that a progenitor gas fraction of 30 per
cent results in remnants parameters in good agreement with the
Fundamental Plane for elliptical galaxies. One interesting consequence of 
gas infall to the center and subsequent star formation is a break in the 
surface brightness profile with excess light at the center 
\citealp{1996ApJ...464..641M,2006MNRAS.373.1013C}, an effect that is not seen in 
collisionless mergers \citealp{2006MNRAS.369..625N}. For a long time this 
break in the surface density profiles has been considered not to be in agreement 
with the majority of elliptical galaxies. However, recent studies indicates that 
a break in the light profile might be a generic feature of low and intermediate 
mass ellitpticals. For example, Sersic functions provide an excellent fit to 
the profiles of Virgo early-type galaxies outside 10-100pc. 
At the centers there is a transition from a luminosity "deficit" to a luminosity 
"excess" with respect to the Sersic fit moving down the early-type 
luminosity function \citep{2006ApJS..165...57C,2006ApJS..164..334F}. It is still debated 
whether this transition is smooth \citep{2007ApJ...671.1456C} or whether it represents 
a true dichotomy (Kormendy et al. 2007, submitted).

It has been known for some time \citep{1989Natur.340..687H} that galaxy
mergers can, by dynamical instabilities, drive large gas fractions to
the center of the remnants probably  causing a starburst and/or
feeding a super-massive black hole \citep{1996ApJ...464..641M, 
1996ApJ...471..115B,2004A&A...418L..27B,2005A&A...437...69B,2005ApJ...620L..79S}.
Processes like this can be observed directly in nearby ultra-luminous
infrared galaxies which are interacting disc galaxies with mass ratios
in the range of 1:1 to 3:1  
\citep{1998ApJ...498..579G,2001ApJ...563..527G,2006ApJ...638..745D}. 
\citet{2008arXiv0802.0508H} have compared surface density profiles 
of observed nearby  mergers to simulations of gas rich disk mergers. They indeed 
find that the observed "excess" central light can be attributed to a young
stellar component that formed from infalling gas during the merger. 

Using a simple model for gas accretion onto a central super-massive black hole it has 
been argued, based on simulations of binary disc mergers, that gas inflow regulated by 
black hole feedback can naturally explain the observed present day relation between 
stellar velocity dispersion and black hole mass for elliptical galaxies and their evolution 
with redshift \citep{2005Natur.433..604D,2006ApJ...641...90R}. Surprisingly, using the 
same model repeated isolated mergers of more evolved systems appear to stay on the 
stellar velocity dispersion and black hole mass relation \citep{2008arXiv0802.0210J}.  
Extended models have 
also been used to understand the evolution of Quasars and stellar spheroids as a whole 
\citep{2005ApJ...630..705H,2006ApJS..163....1H}. However, recent
studies of dissipative mergers by \citet{2006MNRAS.372..839N}, and,
including star formation and feedback processes, by
\citet{2006ApJ...650..791C,2006MNRAS.373.1013C} confirm and strengthen the
\citet{2003ApJ...597..893N} conclusion that binary disk mergers only are
reasonable progenitors of intermediate mass giant ellitpicals but not of the more 
massive ellipticals. These studies are based on very detailed photometric and 
kinematic comparisons.

\section{The Problem} 
\label{PROBLEM}

Despite the success of the binary merger scenario for the formation of elliptical 
galaxies there remain fundamental problems comparing the stellar
populations of present day elliptical and spiral galaxies. These problems concern 
even typical $M_*$ ellipticals which would from a kinematical point of view be consistent 
with the binary disk merger scenario. Several of
the arguments presented in the following were outlined in primitive
form by \citet{1980ComAp...8..177O}.  
  
One aspect concerns the mass in metals of the two kinds of
systems. We have estimated the typical mass of late-type and
early-type galaxies by computing the mass above which half the total 
mass in galaxies of each type is contained assuming a lower limit of $10^9
M_{\odot}$. Using averages of the color and concentration
selected  mass functions of \citet{2003ApJS..149..289B} we
get $M_{\mathrm{early}} = 7.5 \times 10^{10}M_{\odot}$ and $M_{\mathrm{late}} =
2.9 \times 10^{10}M_{\odot}$. With typical metallicities for the
respective populations of $z_{\mathrm{early}}=0.03$ and
$z_{\mathrm{late}}=0.016$  \citep{2006ApJ...653..881N} we get total masses
in metals of $M_{\mathrm{z,early}}=2.3\times 10^9 M_{\odot}$ and 
$M_{\mathrm{z,late}}=4.6\times 10^8 M_{\odot}$. E.g.the mass in
metals is $\approx 4.8$ times higher in typical ellipticals than in
typical discs. Using an alternative approach using $L_*$ for early
type and late type galaxies based on the SDSS data of
\citep{2003AJ....125.1682N} we obtain $M_{*,\mathrm{early}}=1.0\times
10^{11} M_{\odot}$ and $M_{*,\mathrm{late}}=2.5\times 10^{10}
M_{\odot}$  using $M/L_{\mathrm{early}} = 3.19$ and  
$M/L_{\mathrm{early}} = 1.18$ \citep{2006ApJ...653..881N}. With the
same typical metallicities we then get  $M_{\m{*,z,early}}=3.1\times
10^9 M_{\odot}$ and  $M_{\m{*,z,late}}=4.1\times 10^8 M_{\odot}$. This
results in a $\approx 7.7$ times higher mass in metals of ellipticals.   

Following the model for the evolution of the disc of our own galaxy,
the Milky Way, presented in \citet{2006MNRAS.366..899N} and extended
here we find that the stellar component (disc and bulge) has a total
mass of $5\times 10^{10} M_{\odot}$ containing $\approx 8.4 \times 10^8
M_{\odot}$ solar masses 
in metals. The mean metallicity is in good agreement with typical
estimates for the local disk galaxy population as whole, however, the
total mass is slightly above the typical disc mass estimated
before as the Milky Way is more massive than a typical disk galaxy. 
Still we can assume that the evolution with redshift
is similar for all disks. With plausible assumptions for the evolution of the star
formation rate the mass in metals for one disk galaxy was $\approx 50$ percent smaller at
$z=1$ and $\approx 80$ percent smaller at $z=2$. Under the simplified
assumption that early-type galaxies evolve passively since z=2 the
progenitors of a typical present day disc have about 9-15 times less
metals at z=1 and about 20-35 times less at z=2. Simply put the mass
in metals of two typical spirals is significantly less than in typical
ellipticals. And the problem becomes worse as one considers mergers at
z=1 or even z=2. Clearly this problem is most severe for massive
ellipticals.   

Another aspect concerns the ages of the stellar populations. There is
strong observational evidence that evolved, massive, red and 
metal rich proto-ellipticals are already in place at $z=2-3$ and that
present day early-type galaxies formed most of their stars well before a 
redshift $z=1$ \citep{1973ApJ...179..427S,1996MNRAS.281..985V,2000ApJ...536L..77B,1998ApJ...504L..17V,1998ApJ...493..529B,2005ApJ...633..174T,2005ApJ...631..145V,2006ApJ...638L..59V,2006ApJ...649L..71K,2006MNRAS.365.1114P}. For example, 
\citet{2006ApJ...649L..71K} 
found, using near-infrared spectroscopy, that about 45\% of massive galaxies at $z~2$ little/no 
star formation and evolved underlying stellar populations. 
Suprisingly, these galaxies are significantly more compact than present day ellipticals
with equivalent stellar masses \citep{2008ApJ...677L...5V,2008arXiv0801.1184C}.

In general, elliptical galaxies are red and the  
formation of their stars was complete 8-10 Gyrs ago whereas disk galaxies
in general are blue and have much younger  
stellar populations with e.g. mean ages of 5 Gyrs for the Milky Way 
\citep{2003A&A...409..523R}. In combination with the fact that disc galaxies are less 
massive and less concentrated than massive elliptical galaxies it can be excluded that all 
elliptical galaxies could have formed from binary mergers of present day spiral galaxies.  
Of course elliptical galaxies might have formed from the progenitors of present day disc 
galaxies. But observations of the size and mass evolution of disc galaxies 
\citep{2005ApJ...635..959B,2005ApJ...630L..17T,2006ApJ...650...18T} as well as a plethora of 
theoretical models \citep{1997ApJ...477..765C,1998MNRAS.297L..71M,
1998ApJ...507..229P,2000MNRAS.312..398B,2006MNRAS.366..899N} indicate that spiral galaxies 
in the past were in general smaller and less massive than today. As disk merger 
remnants typically have similar projected sizes and not more 
than two times the progenitor mass in the spheroidal component \citep{2006MNRAS.369..625N} 
it is even more unlikely that massive elliptical galaxies could have
formed from binary disc mergers in the past.       

Furthermore it has been shown that the stellar populations of massive
ellipticals have not only formed at high redshift but also on short
timescales
(e.g. \citealp{2004Natur.428..625H,2005ApJ...621..673T,2006MNRAS.365.1114P}). 
This is not compatible with the long formation timescales of disc galaxies. The
problem is less severe for low mass ellipticals which have more
extended formation timescales and even show signs of young stellar
populations \citep{2005ApJ...621..673T}. 

An additional complication is that elliptical galaxies predominantly populate high 
density regions like galaxy clusters \citep{1931ApJ....74...43H,1937C&T....53..194H}, whereas 
disk galaxies populate the field and the overdensity is independent of 
their luminosity at a given color \citep{2003ApJ...585L...5H}. The 
morphology-density relationship seems to be valid over  several magnitudes in density
\citep{1977ApJ...215..401M,1980ApJ...236..351D,1984ApJ...281...95P,1991ApJ...367...64W}
and the most massive ellipticals live in the highest density 
environments \citep{2003ApJ...585L...5H}. In summary, the mass of spiral 
galaxies is a very weak function of environment, early type galaxies 
are predominantly found in high density regions and the more massive the 
galaxy the higher the overdensity. With the exception of accretion onto the 
central cluster galaxy, mergers are not likely in clusters, therefore
in the simplest form of the merger scenario a typical 
$10^{12} M_{\odot}$ elliptical must have formed from a merger of two 
$5 \times 10^{11} M_{\odot}$ disc galaxies outside of the cluster, and
the hypothesized progenitor spirals are rare or nonexistent at any
observed epoch. 

In the following we present a model for the evolution of a typical
spiral galaxy with a central bulge component and address several of
the above questions using an idealised scenario of a binary merger of
our model galaxy and its progenitors. 

\section{The model for bulge and disk formation} 
In this section we describe the simple \citet{2006MNRAS.366..899N}
model for the formation of the Milky Way disc where we added to our
previous treatment the formation of a luminous central bulge component. 
The model reproduces nicely observable properties of the present day 
Milky Way without the explicit inclusion of detailed feedback
processes (e.g.\citealp{2000MNRAS.317..697E}) which have their main 
impact at the early phases of disk evolution \citet{2006MNRAS.371.1519J}.   
We assume that in the absence of star formation the gas in a given halo would 
settle in a disc with an exponential surface density 
\begin{equation}
\Sigma_{\m{d}}(r,t)=\Sigma_0(t) \exp[-r/r_{\m{d}}(t) \label{Sigma}],
\end{equation}
where the central surface density $\Sigma_0(t)$ and the scale length
$r_d(t)$ can change with time. The galactic disk starts to 
form after the halo has reached its present day virial velocity at its 
formation time $t_{\m form}$. Thereafter it
decouples from the general merging. After the decoupling the central 
surface density is fixed to the present day value while the scale 
length $r_d(t)$ is allowed to change as a fixed fraction $f_{\m r,d}$ of
the virial radius $r_{\m{vir}}$ of the halo
\begin{eqnarray}
r_{\m{d}}(t\ge t_{\m{form}})& = &
f_{\m{r,d}}\frac{v_{\m{vir}}(t_{\m{form}})}{10 H(t)},
\end{eqnarray} 
where
\begin{equation} 
H[z(t)]=H_0[\Omega_{\Lambda,0}+(1-\Omega_{\Lambda,0}-\Omega_0)(1+z)^2+\Omega_0(1+z)^3]^{1/2},
\end{equation}
is the Hubble parameter at redshift $z$.

At earlier times  $t < t_{\m form}$ when the galaxy is still coupled to the 
hierarchical growth the gas will have lost its angular momentum
more effectively due to shocks and tidal torques resulting in a 
smaller collapse fraction $f_{\m r,b}$. We associate this phase with 
the formation epoch of the bulge. For simplicity we assume that the gas 
also settles in an exponential surface density profile. The spheroidal stellar 
component might form from dynamical instabilities or tidal interactions which we do not 
follow explicitly in our model. The scale length of the gas then evolves as  
\begin{equation}
r_{\m{b}}(t < t_{\m{form}}) =  f_{\m{r,b}} r_{\m{vir}}(t) = f_{\m{r,b}}
\frac{v_{\m{vir}}(t)}{10 H(t)}
\end{equation}
and rotation velocity of the 'bulge' peaks at a radius of $r_{\m{2.2}}(t)
= 2.15 r_{\m{b}}(t)$ at a value of   
\begin{equation}
v^2_{\m{b,2.2}}(t)= 0.774\pi G \Sigma_0(t) r_{\m{b}}(t) \label{v_d,m}.
\end{equation}

\begin{figure}
\begin{center}
  \epsfig{file=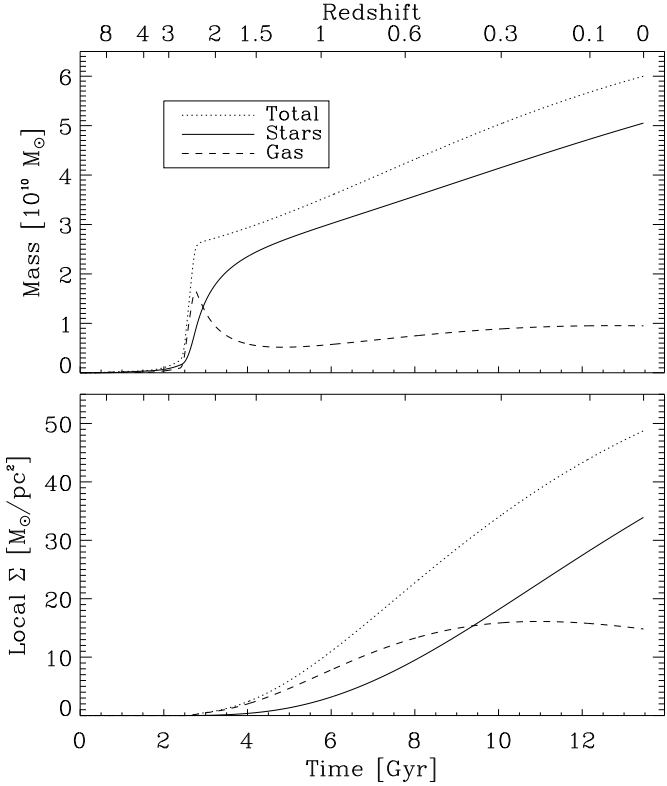, width=0.5\textwidth}
  \caption{{\it{Upper panel}}: Time evolution of the total mass, the
  mass in stars and the mass in gas for the model disc. {\it{Lower
  panel}}: Time evolution of the total surface density, the stellar
  surface density and the gas surface density at the solar
  radius. \label{p_all_mass_vs_time}}   
\end{center}
\end{figure}

\begin{figure}
\begin{center}
  \epsfig{file=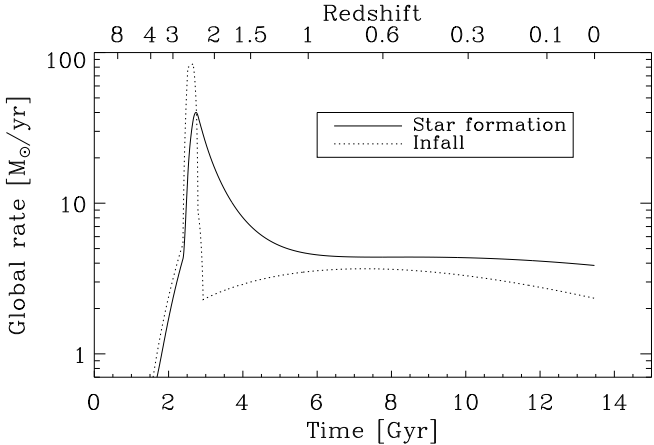, width=0.5\textwidth}
  \caption{Evolution of the global star formation
  rate (solid) and the global infall rate (dotted). The rapid
  formation of the bulge at $3 < z <2$ is followed by the quiescent
  assembly of the disk. \label{p_sfr_vs_time}}   
\end{center}
\end{figure}

Using the similar scalings as in \citet{2006MNRAS.366..899N} the central surface 
density will increase as 
\begin{eqnarray}
\Sigma_0(t <t_{\m{form}})  =  \frac{\alpha}{\pi G}  10 H(t) v_{\m{c}}(t), \label{sig_0_1}
\end{eqnarray}
with $\alpha = 3.6$ that has been scaled to result in a peak rotation
velocity the Milky Way bulge of $250 km/s$ assuming a present day value of
$v_{\m c} = 210 km/s$ at large radii. The total surface density of the
system has been scaled to a present day value of $50M_\odot pc^{-2}$
at $r = 8kpc$,  

At this point we have to include a model for star formation to the
model to follow the evolution of the stellar and gaseous phase
separately. After the gas within a halo has settled it
starts forming stars. We use a formulation based on the local dynamical
time (rotation period) of the system \citep{1998ApJ...498..541K}. At
every radius the surface density of the star formation rate is given
by   
\begin{equation}
\Sigma_{\m{SFR}}(r,t) = \epsilon
\frac{\Sigma_{\m{gas}}(r,t)}{\tau_{\m{orb}}(r,t)}\label{sigma_sfr}  
\end{equation} 
with 
\begin{equation}
\tau_{\m{orb}}(r,t) = \frac{2 \pi r}{v_{\m{c}}(t)}
\end{equation} 
where $\tau_{\m{orb}}$ is the orbital period and the star formation
efficiency is $\epsilon = 0.1$. 

We follow the chemical evolution of the model galaxies in independent rings
assuming no radial gas flows using a modified version of the chemical
evolution model of \citet{1975ApJ...201L..51O}. In every ring the
change in gas surface density  $\Sigma_{\m{g}}$  and surface 
density in stars $\Sigma_{\m{s}}$ is given by   
\begin{eqnarray}
d\Sigma_{\m{g}}(r,t) &=&  -\Sigma_{\m{SFR}}(r,t)dt
+K_{\m{ins}}(r,t)dt \nonumber \\
& & +K_{\m{late}}(r,t)dt +\Sigma_{\m{IFR}}(r,t)dt \label{dmg}  \\ 
d\Sigma_{\m{s}}(r,t) &=& \,\,\,\, \Sigma_{\m{SFR}}(r,t)dt- K_{\m{ins}}(r,t)dt \nonumber \\
& & -K_{\m{late}}(r,t)dt   \label{dms},
\end{eqnarray}
where $\Sigma_{\m{SFR}}$ is the star formation rate per unit area
(Eqn. \ref{sigma_sfr}) and $\Sigma_{\m{IFR}}$ is the rate of gas
infall onto the galaxy per unit area, as defined in
Eqn. \ref{sigma_sfr}. $K_{\m{ins}}$ is the mass per unit are in gas 
ejected from massive stars instantaneously, $K_{\m{late}}$ is the mass
per unit area in gas ejected at later evolutionary phases of low mass
stars. They are defined as  
\begin{eqnarray}
K_{\m{ins}}(r,t) & = & R_{\m{ins}} \Sigma_{\m{SFR}}(r,t),\label{K_1} \\
K_{\m{late}}(r,t) & = &\int_0^t \Sigma_{\m{SFR}}(t^\prime,r) W(t-t^\prime) dt^\prime.\label{K_2}
\end{eqnarray}
$R_{\m{ins}}=0.1$ is the fraction of gas returned instantaneously to the
ISM from newborn massive stars and $W(t)$ is a weighting function
defined as    
\begin{equation}
W(t) = R_* \frac{\delta_*-1}{\tau_0}
\left(\frac{\tau_0}{t+\tau_0}\right)^{\delta_*}  \label{ciotti} 
\end{equation}
with $R_* = 0.3$, $\delta_* = 1.36$, and $\tau_0 = 1 \times 10^8$ assuming a \citet{1955ApJ...121..161S} IMF
\citep{1991ApJ...376..380C}. This analytic expression is a good
approximation for the fraction of
returned gas for a single burst population to the metal dependent
values of the spectral evolution model by \citet{2003MNRAS.344.1000B}
that we have used to compute the photometric properties of the model
disc not including the effects of dust (see \citealp{2006MNRAS.366..899N}). We followed the evolution of the 
metallicity in gas and stars both taking into account instantaneous injection from 
massive stars and delayed injection form low mass stars. The details of the chemical evolution model 
are given in \citet{2006MNRAS.366..899N}. 

\subsection{Assembly history}

The upper panel in Fig. \ref{p_all_mass_vs_time} shows the evolution
of the total mass in gas and stars, respectively. The galaxy rapidly
assembles mass after $z=3$, which is the bulge formation phase,
thereafter the disk is growing. At present the total mass
of the disc out to $26 kpc$ is $M_{\m{tot}}\approx 6\times 10^{10}
M_{\m{\odot}}$  with about $M_{\m{g}}=1\times 10^{10} M_{\odot}$ in gas and
$M_{\m{tot}}= 5\times 10^{10} M_{\odot}$ in stars. 

\begin{figure}
\begin{center}
  \epsfig{file=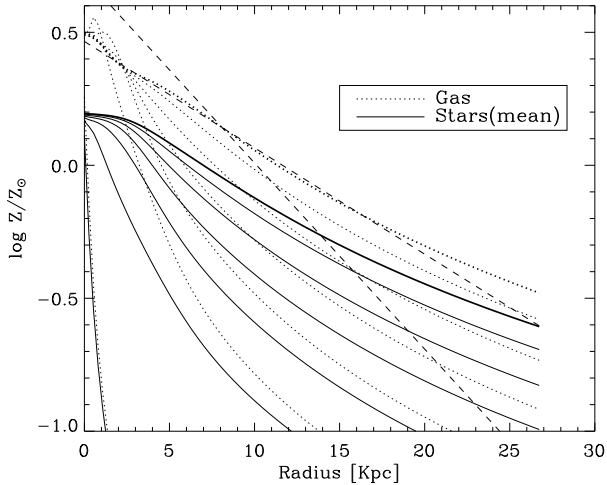, width=0.5\textwidth}
  \caption{Radial metallicity distribution of the gas (solid lines)
  and the mean metallicity of the stars (dotted lines) after 2, 4, 6,
  8, 10, 12  and 13.6 $Gyrs$ (thick lines). The two dotted straight
  lines indicate a slope of   $-0.07 dex kpc^{-1}$ and $-0.04 dex
  kpc^{-1}$. The metallicity gradients have been steeper in the
  past.\label{p_z_vs_rad}} 
\end{center}
\end{figure}

The time evolution of the local surface density is shown in the lower
plot in Fig. \ref{p_all_mass_vs_time}. The model has been normalised
to a present day total surface density of $50 M_{\odot} pc^{-2} $. The
gas starts to assemble at the solar radius after $3 Gyrs$, and after $6
Gyrs$ the local gas surface density stays almost constant at its present day
value of  $15 M_{\odot} pc^{-2}$.  The first stars at the solar radius
form after $4 Gyrs$ followed by a steady increase to the present day value
of $35 M_{\odot} pc^{-2}$ (with $\approx 3 M_{\odot} pc^{-2}$ invisible in
stellar remnants) resulting in $\approx 32 M_{\odot} pc^{-2}$ visible stars. 
These numbers are identical to those given in
\citet{2006MNRAS.366..899N} as is the local 
star formation history at the solar radius. Therefore the
evolution of the outer disk is not affected by the bulge evolution. 
\begin{figure}
\begin{center}
  \epsfig{file=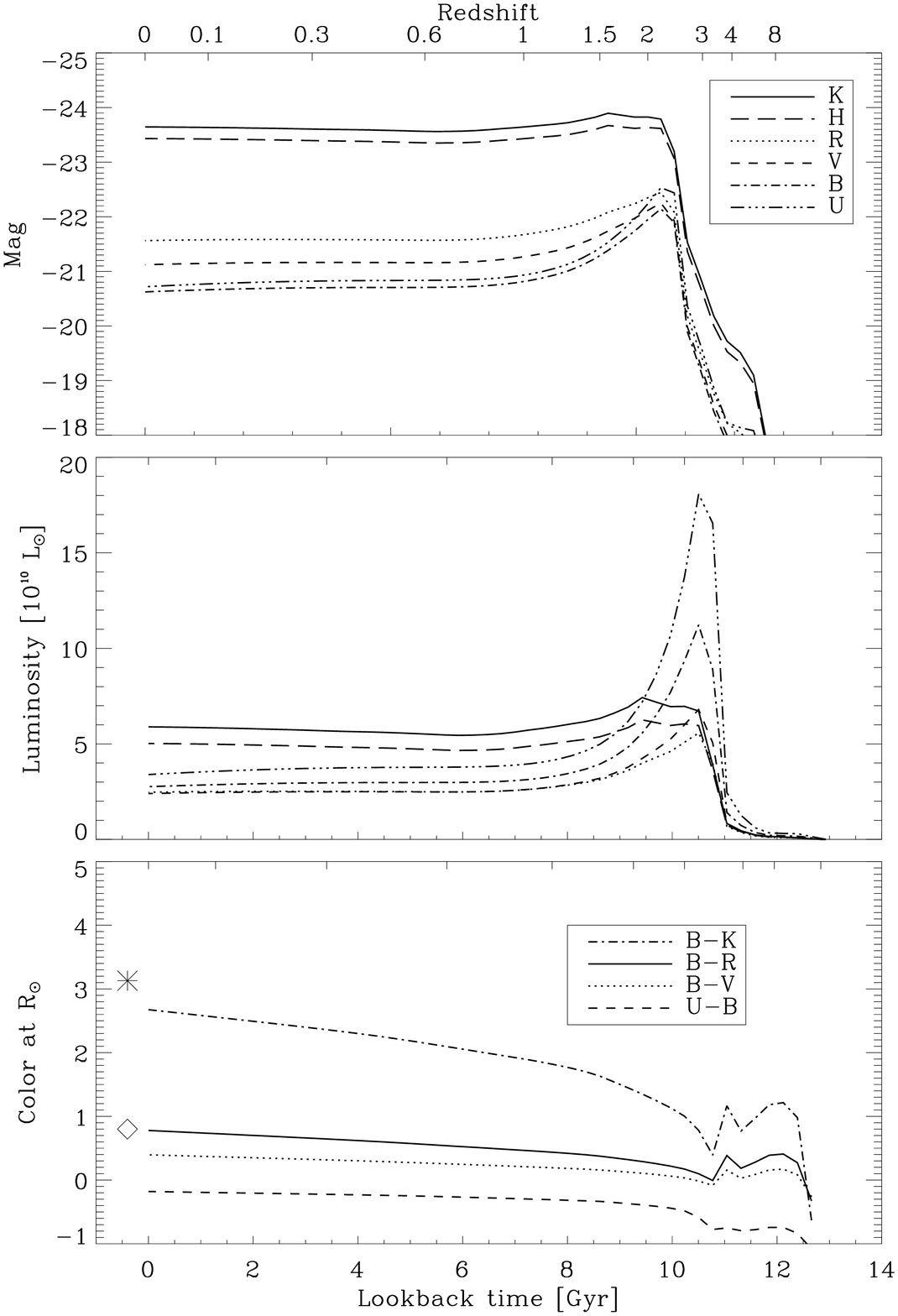, width=0.5\textwidth}
  \caption{{\it{Upper panel}}: Absolute magnitude of the model galaxy
  versus lookback time.{{\it{Middle panel}}: Evolution of the total
  luminosities with time.{\it Bottom panel}: Evolution of the
  color at the solar radius. The observed colors are $B-K = 3.13$
  \citep{1998gaas.book.....B} indicated by the star and $B-V \approx
  0.8$ \citep{1999MNRAS.307..857B} indicated by the diamond. 
   \label{p_all_photo_vs_time}} }
\end{center}
\end{figure}

The evolution of the total star formation and infall rate of
the model galaxy is shown in Fig. \ref{p_sfr_vs_time}
The peak in both rates at $3<z<2$ indicates the period of the
formation of the bulge with infall rates as high as 
$90 M_{\odot}  yr^{-1}$ and star formation rates of $\approx 30
M_{\odot}  yr^{-1}$. It is interesting that this analytical treatment
of the formation of the galactic bulge parallels to a surprising
degree the detailed hydrodynamical simulations of an elliptical galaxy
by \citet{2007ApJ...658..710N} from cosmological initial
conditions. Thereafter the rates drop rapidly to values close 
to their present day numbers of $2-3 M_{\odot}  yr^{-1}$ for the
infall rate and $\approx 3 M_{\odot}  yr^{-1}$ for the star formation
rate.   

As mentioned before, the evolution of the model galaxy at the
solar radius is not changed by the inclusion of the bulge. Therefore
the metallicity distribution is the same and in \citet{2006MNRAS.366..899N} and is in good
agreement with observations. To compute the metallicity evolution we
have used the solar metallicity value of $Z_{\odot}=
0.0126$ to scale our model to 0.1 dex below the solar metallicity $Z =
Z_{\odot}\times  10^{-0.1} = 0.01$ at solar radius $4.5 Gyrs$ ago,
resulting in an effective yield of $Y=0.0135$ (see \citealp{2006MNRAS.366..899N}). 

Fig. \ref{p_z_vs_rad} shows the radial metallicity distribution of the
ISM and the stars of the model galaxy. The distribution after $2, 4,
6, 8, 10,$ and $12 Gyrs$ is indicated by the dotted (gas) and solid
(stars) lines. At the solar radius the metallicity gradient is similar
to the pure disc model of \citet{2006MNRAS.366..899N}, however, at smaller radii the
metallicity of the gas and the stars is significantly higher.

\subsection{Photometric properties}
\label{photo}

To compute the photometric properties of our model galaxy we use the
\citet{2003MNRAS.344.1000B} models for spectral evolution of stellar
populations assuming a Salpeter IMF. Fig.\ref{p_all_photo_vs_time}
shows the time evolution of the absolute magnitudes, the total
luminosities and the colours of the model galaxy at different
wavelengths. All values are in reasonable agreement with observations.  
\begin{figure}
\begin{center}
  \epsfig{file=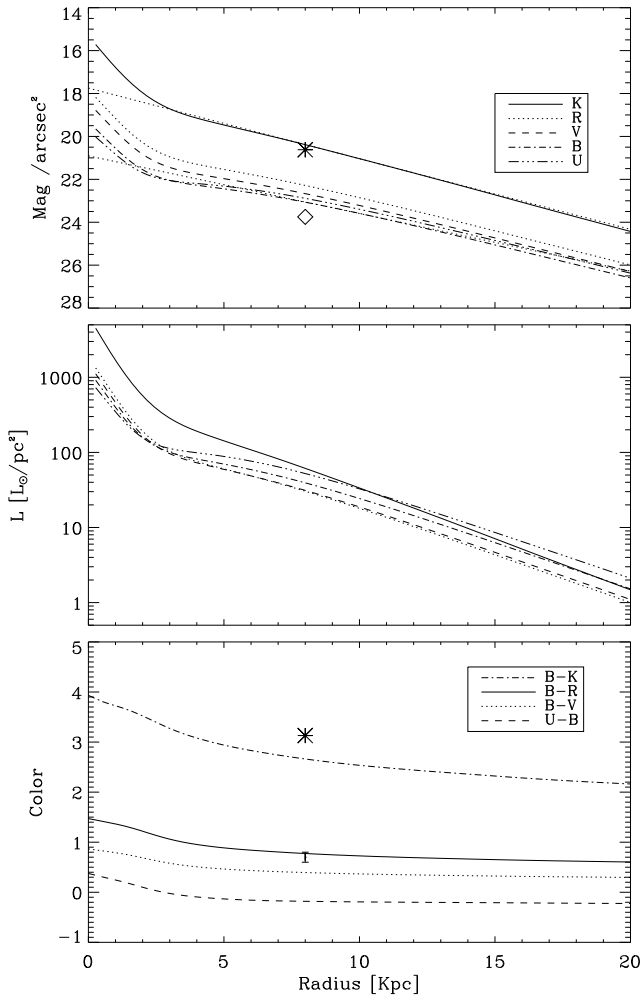, width=0.5\textwidth}
  \caption{Photometric properties versus radius.
	Observed values are $\mu_{\m{K}}=20.62$ (star) and 
        $\mu_{\m{B}} = 23.75$ (diamond)	\citep{1998gaas.book.....B} 
	The observed colors are $B-K = 3.13$ \citep{1998gaas.book.....B} indicated by 
the star in the lowest panel and $0.6 < B-V < 0.8$ (see \citealp{1999MNRAS.307..857B}) indicated by the error bar. \label{p_all_photo_vs_rad}}
\end{center}
\end{figure}
The surface brightness profiles at different wavelengths are shown in
the upper panel of Fig. \ref{p_all_photo_vs_rad}. Two observed values
in the K- and B-band at the solar radius are overplotted
\citep{1998gaas.book.....B}. The middle panel shows the corresponding
luminosity density profiles using the solar magnitudes given
above. All profiles are exponential in the outer parts. The color
profiles with two observational value at the solar radius are shown
in the bottom panel of Fig. \ref{p_all_photo_vs_rad}. We have performed
a bulge-disc decomposition using the fitting model described in
\citet{2006MNRAS.369..625N} and find a bulge to total ratio of
$B/T$=0.2 in reasonable agreement with estimates for the Milky Way. 

\begin{figure}
\begin{center}
  \epsfig{file=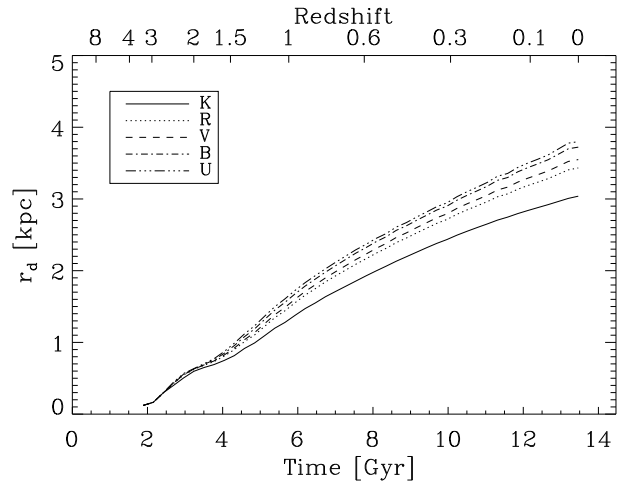, width=0.5\textwidth}
  \caption{Exponential scale lengths of the surface brightness
  distribution measured at the different wavelengths versus time. At
  shorter wavelengths the disc has a larger scale length. 
 \label{p_ubv_scale_vs_time}}
\end{center}
\end{figure}

To calculate the scale lengths in the different bands we have fitted an  
exponential in the range of $1.5 < r_{\m{d}} < 3.0$ scale lengths of
the total surface density distribution. The scale lengths increase to
shorter wavelengths (Fig. \ref{p_ubv_scale_vs_time}). This effect is
weaker at earlier times. At present the B-band scale length
$r_{\m{d,B}} = 3.7 kpc$ is a factor of $\approx 1.2$ larger than the
K-band scale length $r_{\m{d,K}} = 3.0 kpc$ . This trend is observed
and is in good agreement with other Milky Way type spiral
galaxies. \citet{1996A&A...313..377D} investigated a sample of 86
nearly face-on spiral galaxies and concluded that spiral galaxies of
the same type as the Milky Way have disc scale lengths that are a
factor of $1.25 \pm 0.25$ larger in the B-band than in the
K-band. This effect reflects the over all inside out formation process
for disc galaxies and has been found in similar studies
(\citet{1999MNRAS.307..857B} have found  $r_{\m{d,B}}/r_{\m{d,K}} =
1.5$ which is also in good agreement). We have separately examined the
age distribution of the stars in the model Milky Way galaxy and find
them in good agreement with observations, giving additional credence
to our believes that the backward evolution of the model galaxy is
plausible. To summarize the evolution of our model Milky Way, it was
smaller and less massive in the past but since $z\approx 2$ its
luminosity and metallicity have been remarkably constant.    

\section{The idealised disc merger} 
\label{CONSTRAINTS}

\begin{figure}
\begin{center}
  \epsfig{file=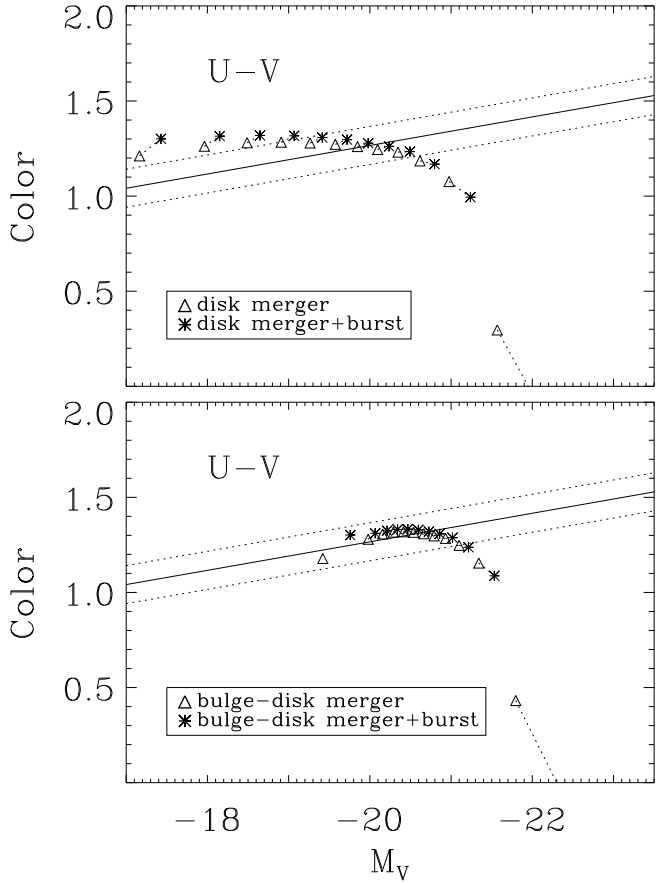, width=.5\textwidth}
  \caption{{\it Upper panel:} $U-V$ color versus absolute magnitude
    $M_V$ of the stellar populations of two pure disk galaxies added
    together in an idealized merger at different times in the
    past. The triangles show the properties for the stellar remnant
    alone. The asterixes show the remnant properties assuming that all
    gas within two scale lengths at the given time was transformed
    into stars in a single burst with a duration of 150 Myrs. The burst and no-burst models are 
    connected by a dotted line. The present day burst model is off the
    scale. The rightmost pair indicates
    the properties of a present day merger remnant. Every pair
    further to the left shifts the merger event one Gyr further back
    into the past.  {\it Lower panel:} Same as above but for for
    progenitor disks with bulges.  All remnants are consistent with
    the CM-relation if the merger would have taken 
    place $3$ Gyrs ago or earlier. $5$ Gyrs or older mergers of disk
    galaxies with bulges lead to the most consistent results. Mergers
    of present day disks  do not fall on the CM-relations independent
    of their properties.\label{p_evol_uv_vs_mag} } 
\end{center}
\end{figure}

Numerical studies in the past have shown that sizes, photometric
properties and dynamics of disk merger remnants are in reasonable
agreement with the properties of observed intermediate mass giant
elliptical galaxies. Using the model for galaxy evolution described in 
the previous section we now have a self consistent model giving
knowledge about the evolution 
of a typical disc galaxy, e.g. ages and metallicities of the stars and
gas content at any redshift. Using this
information we can investigate the remnant properties in an idealised
disk galaxy merger scenario. We assume that two identical disk
galaxies (with or without bulges) merge at different times in the
past and investigate the integrated properties of their combined
stellar populations today. During the merger we allow all gas within
two disc scale lengths to be transformed into stars in a single
burst. The duration of the burst was taken from typical burst timescales of 
equal mass binary disk merger simulations (see e.g. \citealp{2008arXiv0802.0210J}).
Somewhat smaller or larger durations do not influence the results.  
After the galaxies have merged we assume that further star
formation is suppressed. This might either be due to the heating of
gas by supernovae, AGN feedback, shock heating of infalling gas, the
infall of the remnant into a cluster or a combination of all
effects. As we know the properties of the stars and the gas content of
the progenitors disks at any redshift, we can predict the present day
properties of the stellar population of the remnant that has formed by
a merger at any time in the past.

In Figure \ref{p_evol_uv_vs_mag} we show the present day location
of the merger remnants of two Milky Way like galaxies in the $U-V$
color-magnitude diagram. For every merger model the rightmost symbols  
represents the properties of the present day merger. Following the
symbols for every model from the right to the left we have placed the
merger further back in the past. The observed UV color-magnitude diagram
of elliptical galaxies is overplotted \citep{2005ApJ...619..193M}. In
the two upper panels we show the properties of a merger of pure disks
with and without a burst assuming the \citet{2006MNRAS.366..899N}
model.  It is evident that a present day merger of disk galaxies
results in a remnant that is far 
too blue to be an elliptical galaxy. The discrepancy is even stronger
if we consider the young population of stars created in a burst during
the merger. Placing the merger further back in time results in a
present day remnant that becomes less luminous and redder. Taking the
spread in the CM-relation into account the remnant properties would
be consistent with both CM-relations if the galaxies would have merged
$3-8$ Gyrs ago. The remnant would fall right on the CM-relations if
the merger would have taken place $5$ Gyrs ago and would give rise to
an elliptical galaxy with an absolute V-band magnitude of $M_V
=-19.8$ at present. Surprisingly, if we go even further back in time the remnant
of the model without a bulge would lie above $U-V$ CM-relation leaving
two possible interpretations: either the remnant is too red or it is
simply not luminous enough.

\begin{figure}  
\begin{center}
  \epsfig{file=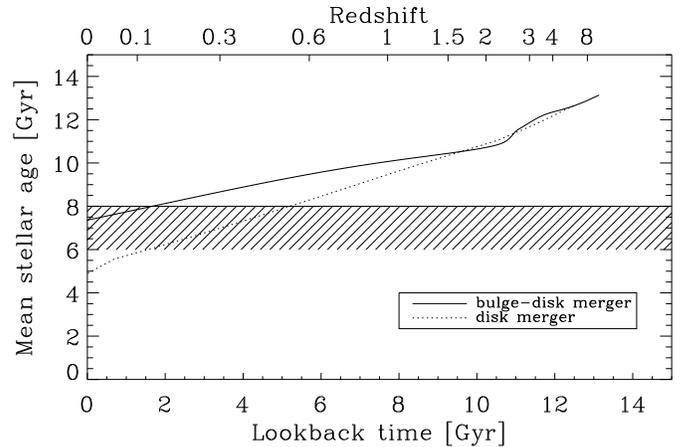, width=.5\textwidth}
  \caption{Mean age of the stellar population of disk mergers with and
    without bulges versus the lookback time of the merger. The
    horizontal line indicates a lower limit of $8$ Gyrs for the age of
    giant elliptical galaxies. All younger ages (shaded area) are not
    compatible with the stellar populations of elliptical
    galaxies. \label{p_age_vs_time}} 
\end{center}
\end{figure}

If we add a bulge to the progenitor disk using the model presented in
the previous section the remnants become redder and more luminous at
all times (lower panel in Fig. \ref{p_evol_uv_vs_mag}). As in
the disk-only case we have to place the merger event at least $3$ Gyrs
in the past to be consistent  with the present day UV CM-relation. The
$5$ Gyr old merger falls right on the correlation as in the pure disk
case and produces an elliptical having a current magnitude of $M_V
=-20.6$. In contrast to the disk-only models, all earlier mergers fall on
the color-magnitude relations due to the presence of the massive old
bulge component. However, at higher redshifts $z>1$ it is  a merger of
two bulge dominated galaxies and not a disc merger, i.e. to a
significant extent we are considering the merger of ellipticals to
make new elliptical.

In Fig. \ref{p_age_vs_time} we show the mean age of the stellar
population of the merger remnant at different lookback times for the
models with and without a bulge component. For a $5$ Gyr old merger
(which corresponds to a merger redshift of $z \approx 0.5$) the
disk-only remnant has a mean age of $7.9$ Gyrs. The remnant with
bulges in the progenitor disks is significantly older at a
mean age of $9.2$ Gyrs.  
                                   
\begin{figure}  
\begin{center}
  \epsfig{file=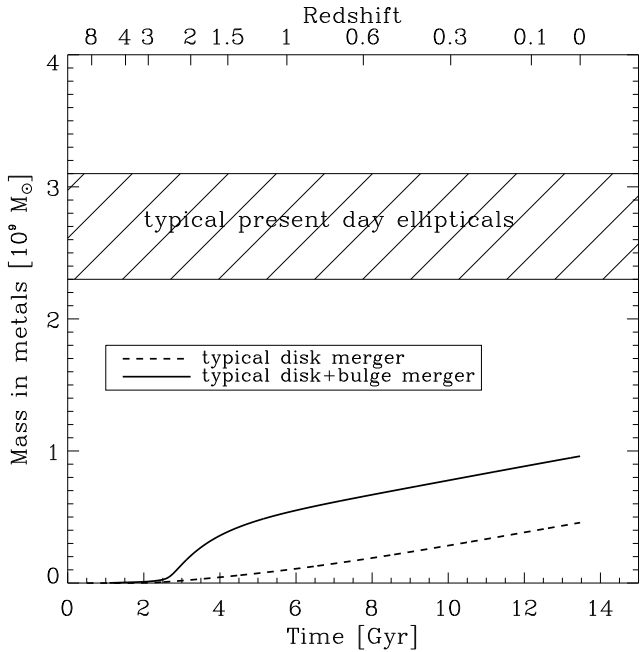, width=.5\textwidth}
  \caption{Evolution of the mass in metals for merger of two  $M_*$ disk galaxies based on 
the model presented in the paper with (solid) and without (dashed) a
bulge component.  
The total mass of the progenitor disks has been scaled to $2.9 \times 10^{10} M_{\odot}$.
The mass in metals for $M_*$ ellipticals is indicated by the shaded
area (see Sec. \ref{PROBLEM}). At present ellipticals have at least a factor 
of two more mass in metals.     
\label{p_mass_metals_vs_time}}
\end{center}
\end{figure}

Our model Milky Way galaxy with bulge is about two times more massive than
a typical present day disc galaxy. If we, however, assume that the
backward evolution with time is typical for disk galaxies of similar
mass, and there is observational evidence for that
\citep{2005ApJ...635..959B}, we can scale our model down and follow
the evolution of the mass in metals for a typical disk merger backwards in
time. In Fig. \ref{p_mass_metals_vs_time} we have scaled the model to
a present day total stellar mass of $2.9 \times 10^{10} M_{\odot}$
(see Section \ref{PROBLEM}) and compare the mass in metals  
of disc mergers with and without bulge to present day ellipticals. At
present, disc merger remnants have $\approx 2.4 - 3.1$ times less mass in
metals if they contain bulge components. Without bulges this value
becomes as large as 6-8 at the present day and even larger at higher redshifts.   
The bulge, although less massive than the disk itself is significantly
more metal rich, in line with our previous arguments. Therefore the total
mass in metals is about a factor of two higher for the model with bulge.
Assuming a burst during the merger does not change the overall values
much as the total gas fraction of the progenitor galaxies stays at a
constant fraction of $\approx 20\%$ since z=2. The metal mass of
ellipticals could only be reached if all available gas would be
transformed into stars with eight times the solar metallicity during
the merger.

\section{Conclusion and Discussion}
In this paper we have addressed the question of whether elliptical
galaxies can have formed from binary mergers of disc galaxies. 
At present typical disc galaxies have lower stellar masses, lower
masses in metals, younger stellar populations and more extended star
formation histories than elliptical galaxies. In addition they
populate different environments. The first simple conclusion is that mergers
of present day spiral galaxies, like our Milky Way, can not form
typical present day ellipticals but some subset of future ellipticals having
$L \leq L_*$ may form in this fashion right now. In fact nearby
ULIRGS are disc mergers with mass ratios of 1:1 to 3:1
\citep{2006ApJ...638..745D} and they show kinematic and isophotal 
properties similar to present day intermediate mass ellipticals
\citep{2001ApJ...563..527G,2004AJ....128.2098R,2006ApJ...638..745D,2006ApJ...651..835D,2006astro.ph..4493R}
After their stellar populations have aged for $\approx 3-4 Gyrs$ they
will add to the population of intermediate mass ellipticals. Expanding the
investigation to cosmic history we have used a theoretical model for
mergers of a typical Milky Way spiral at different epochs to put more
general constraints on binary merger scenario.   

Assuming that most present day spirals formed in way similar to our model 
presented here we can place further constraints on the formation mechanisms 
of early-type galaxies.  

We find that mergers of progenitors of present day spiral disc
galaxies can have stellar populations comparable to present day low
and intermediate mass elliptical galaxies if they have merged more
than 3-4 Gyrs ago ($z >0.3$) and have their star formation turned off
after the merger event. Any further star formation after the merger will shift
the time for the merger event further into the past. The progenitors
of mergers before z=0.3, however, have smaller sizes ($r \leq 2.5 kpc$), 
smaller stellar masses ($M < 4 \times 10^{10} M_{\odot}$) and are more
bulge dominated than at present. As we know from a large number of
numerical studies (see Section \ref{PROBLEM}) these will have sizes
and kinematical properties similar to intermediate mass ($M_*$)
ellipticals.  And early mergers of spiral galaxies will lead to ellipticals 
having metal masses far below those seen in present day $L_*$ ellipticals.  

Specifically, if the merger takes place further back in the past, say at z=1, the
typical sizes of discs will be ($r \leq 2 kpc$) with stellar masses
$M < 3 \times 10^{10} M_{\odot}$. However, the typical mass in metals
will be a factor of 4-8 smaller than typical present day ellipticals, and,
even worse, at $z=2$ the mass in metals will be a factor of 10-18 smaller
and the merger progenitors will be small and bulge dominated. This
epoch, however, is supposed to be the main formation epoch of
ellipticals, especially the most massive ones.

Again, assuming that most present day disks formed in a similar way, we can conclude 
that typical massive giant ellipticals 
(more massive than $M_*$) can neither be made from binary mergers of present day
spirals nor from their progenitors although some low mass subset (less massive than $M_*$) 
of the total observed z=0 elliptical galaxy population is certainly produced by such
mergers. Alternatively, massive ellipticals could have formed by either multiple mergers
 \citep{2006astro.ph..8190L} or binary mergers of 
massive spirals at early times whose descendents no longer exist. These massive high redshift 
spirals would have had different formation histories than the model galaxy presented here. 
Any scenario
addressing the full mass range of ellipticals that is based on binary
mergers, e.g. for the origin of the black-hole-mass
velocity-dispersion relation (e.g. \citealp{2005Natur.433..604D}, \citealp{2008arXiv0802.0210J}),
will have to account for this conclusion. However, there is clear
evidence that dissipation played an important role during the
formation of the spheroids especially at lower masses (see
e.g.
\citealp{1989ApJ...342L..63K,1992ApJ...399..462B,1996ApJ...464L.119K,2006MNRAS.370.1445D,2006MNRAS.372..839N,2006ApJ...648L..21K,
  2006astro.ph.11328C,2007ApJ...658..710N}).  However, even massive
ellipticals with rotation supported, old and metal rich  kinematically
decoupled cores show clear signs of dissipation
\citep{1992A&A...258..250B,1995A&A...298..405S,2001ApJ...548L..33D,2006MNRAS.373..906M}.

Major binary mergers of early-type galaxies might play an important
role for the final assembly of massive ellipticals 
\citep{2003ApJ...597L.117K,2006ApJ...641...21R,2006MNRAS.369.1081B,2006ApJ...640..241B,2006ApJ...636L..81N,2006ApJ...652..270B}
but they cannot  be connected to the formation of the spheroids in the first
place.  




\begin{acknowledgements}
We thank R. Bender and K. Nagamine for valuable input and P. Johansson for
helpful discussions on the manuscript. We also thank the anonymous referee for 
valuable comments. 
\end{acknowledgements}

\bibliographystyle{apj}
\bibliography{../../REFERENCES/references}

\begin{thebibliography}{142}
\expandafter\ifx\csname natexlab\endcsname\relax\def\natexlab#1{#1}\fi

\bibitem[{{Aceves} {et~al.}(2006){Aceves}, {Vel{\'a}zquez}, \&
  {Cruz}}]{2006MNRAS.373..632A}
{Aceves}, H., {Vel{\'a}zquez}, H., \& {Cruz}, F. 2006, \mnras, 373, 632

\bibitem[{{Barden} {et~al.}(2005){Barden}, {Rix}, {Somerville}, {Bell},
  {H{\"a}u{\ss}ler}, {Peng}, {Borch}, {Beckwith}, {Caldwell}, {Heymans},
  {Jahnke}, {Jogee}, {McIntosh}, {Meisenheimer}, {S{\'a}nchez}, {Wisotzki}, \&
  {Wolf}}]{2005ApJ...635..959B}
{Barden}, M. {et~al.} 2005, \apj, 635, 959

\bibitem[{{Barnes} \& {Hut}(1986)}]{1986Natur.324..446B}
{Barnes}, J., \& {Hut}, P. 1986, \nat, 324, 446

\bibitem[{{Barnes}(1988)}]{1988ApJ...331..699B}
{Barnes}, J.~E. 1988, \apj, 331, 699

\bibitem[{{Barnes}(1990)}]{1990Natur.344..379B}
---. 1990, \nat, 344, 379

\bibitem[{{Barnes}(1992)}]{1992ApJ...393..484B}
---. 1992, \apj, 393, 484

\bibitem[{{Barnes}(1998)}]{1998giis.conf..275B}
{Barnes}, J.~E. 1998, in Saas-Fee Advanced Course 26: Galaxies: Interactions
  and Induced Star Formation, 275

\bibitem[{{Barnes} \& {Hernquist}(1992)}]{1992ARA&A..30..705B}
{Barnes}, J.~E., \& {Hernquist}, L. 1992, \araa, 30, 705

\bibitem[{{Barnes} \& {Hernquist}(1996)}]{1996ApJ...471..115B}
---. 1996, \apj, 471, 115

\bibitem[{{Barnes} \& {Hut}(1989)}]{1989ApJS...70..389B}
{Barnes}, J.~E., \& {Hut}, P. 1989, \apjs, 70, 389

\bibitem[{{Bell} {et~al.}(2003){Bell}, {McIntosh}, {Katz}, \&
  {Weinberg}}]{2003ApJS..149..289B}
{Bell}, E.~F., {McIntosh}, D.~H., {Katz}, N., \& {Weinberg}, M.~D. 2003, \apjs,
  149, 289

\bibitem[{{Bell} {et~al.}(2006{\natexlab{a}}){Bell}, {Naab}, {McIntosh},
  {Somerville}, {Caldwell}, {Barden}, {Wolf}, {Rix}, {Beckwith}, {Borch},
  {H{\"a}ussler}, {Heymans}, {Jahnke}, {Jogee}, {Koposov}, {Meisenheimer},
  {Peng}, {Sanchez}, \& {Wisotzki}}]{2006ApJ...640..241B}
{Bell}, E.~F. {et~al.} 2006{\natexlab{a}}, \apj, 640, 241

\bibitem[{{Bell} {et~al.}(2006{\natexlab{b}}){Bell}, {Phleps}, {Somerville},
  {Wolf}, {Borch}, \& {Meisenheimer}}]{2006ApJ...652..270B}
{Bell}, E.~F., {Phleps}, S., {Somerville}, R.~S., {Wolf}, C., {Borch}, A., \&
  {Meisenheimer}, K. 2006{\natexlab{b}}, \apj, 652, 270

\bibitem[{{Bender}(1988)}]{1988A&A...193L...7B}
{Bender}, R. 1988, \aap, 193, L7

\bibitem[{{Bender} {et~al.}(1992){Bender}, {Burstein}, \&
  {Faber}}]{1992ApJ...399..462B}
{Bender}, R., {Burstein}, D., \& {Faber}, S.~M. 1992, \apj, 399, 462

\bibitem[{{Bender} {et~al.}(1994){Bender}, {Saglia}, \&
  {Gerhard}}]{1994MNRAS.269..785B}
{Bender}, R., {Saglia}, R.~P., \& {Gerhard}, O.~E. 1994, \mnras, 269, 785

\bibitem[{{Bender} {et~al.}(1998){Bender}, {Saglia}, {Ziegler}, {Belloni},
  {Greggio}, {Hopp}, \& {Bruzual}}]{1998ApJ...493..529B}
{Bender}, R., {Saglia}, R.~P., {Ziegler}, B., {Belloni}, P., {Greggio}, L.,
  {Hopp}, U., \& {Bruzual}, G. 1998, \apj, 493, 529

\bibitem[{{Bender} \& {Surma}(1992)}]{1992A&A...258..250B}
{Bender}, R., \& {Surma}, P. 1992, \aap, 258, 250

\bibitem[{{Bender} {et~al.}(1989){Bender}, {Surma}, {Doebereiner},
  {Moellenhoff}, \& {Madejsky}}]{1989A&A...217...35B}
{Bender}, R., {Surma}, P., {Doebereiner}, S., {Moellenhoff}, C., \& {Madejsky},
  R. 1989, \aap, 217, 35

\bibitem[{{Bendo} \& {Barnes}(2000)}]{2000MNRAS.316..315B}
{Bendo}, G.~J., \& {Barnes}, J.~E. 2000, \mnras, 316, 315

\bibitem[{{Binney} \& {Merrifield}(1998)}]{1998gaas.book.....B}
{Binney}, J., \& {Merrifield}, M. 1998, {Galactic astronomy} (Galactic
  astronomy / James Binney and Michael Merrifield.~ Princeton, NJ : Princeton
  University Press, 1998.~ (Princeton series in astrophysics) QB857 .B522 1998
  (\$35.00))

\bibitem[{{Boissier} \& {Prantzos}(1999)}]{1999MNRAS.307..857B}
{Boissier}, S., \& {Prantzos}, N. 1999, \mnras, 307, 857

\bibitem[{{Boissier} \& {Prantzos}(2000)}]{2000MNRAS.312..398B}
---. 2000, \mnras, 312, 398

\bibitem[{{Bournaud} {et~al.}(2004){Bournaud}, {Combes}, \&
  {Jog}}]{2004A&A...418L..27B}
{Bournaud}, F., {Combes}, F., \& {Jog}, C.~J. 2004, \aap, 418, L27

\bibitem[{{Bournaud} {et~al.}(2005){Bournaud}, {Jog}, \&
  {Combes}}]{2005A&A...437...69B}
{Bournaud}, F., {Jog}, C.~J., \& {Combes}, F. 2005, \aap, 437, 69

\bibitem[{{Boylan-Kolchin} {et~al.}(2006){Boylan-Kolchin}, {Ma}, \&
  {Quataert}}]{2006MNRAS.369.1081B}
{Boylan-Kolchin}, M., {Ma}, C.-P., \& {Quataert}, E. 2006, \mnras, 369, 1081

\bibitem[{{Brinchmann} \& {Ellis}(2000)}]{2000ApJ...536L..77B}
{Brinchmann}, J., \& {Ellis}, R.~S. 2000, \apjl, 536, L77

\bibitem[{{Bruzual} \& {Charlot}(2003)}]{2003MNRAS.344.1000B}
{Bruzual}, G., \& {Charlot}, S. 2003, \mnras, 344, 1000

\bibitem[{{Carlberg}(1986)}]{1986ApJ...310..593C}
{Carlberg}, R.~G. 1986, \apj, 310, 593

\bibitem[{{Chiappini} {et~al.}(1997){Chiappini}, {Matteucci}, \&
  {Gratton}}]{1997ApJ...477..765C}
{Chiappini}, C., {Matteucci}, F., \& {Gratton}, R. 1997, \apj, 477, 765

\bibitem[{{Cimatti} {et~al.}(2008){Cimatti}, {Cassata}, {Pozzetti}, {Kurk},
  {Mignoli}, {Renzini}, {Daddi}, {Bolzonella}, {Brusa}, {Rodighiero},
  {Dickinson}, {Franceschini}, {Zamorani}, {Berta}, {Rosati}, \&
  {Halliday}}]{2008arXiv0801.1184C}
{Cimatti}, A. {et~al.} 2008, ArXiv e-prints, 801

\bibitem[{{Ciotti} {et~al.}(2006){Ciotti}, {Lanzoni}, \&
  {Volonteri}}]{2006astro.ph.11328C}
{Ciotti}, L., {Lanzoni}, B., \& {Volonteri}, M. 2006, ArXiv Astrophysics
  e-prints

\bibitem[{{Ciotti} {et~al.}(1991){Ciotti}, {Pellegrini}, {Renzini}, \&
  {D'Ercole}}]{1991ApJ...376..380C}
{Ciotti}, L., {Pellegrini}, S., {Renzini}, A., \& {D'Ercole}, A. 1991, \apj,
  376, 380

\bibitem[{{C{\^o}t{\'e}} {et~al.}(2007){C{\^o}t{\'e}}, {Ferrarese},
  {Jord{\'a}n}, {Blakeslee}, {Chen}, {Infante}, {Merritt}, {Mei}, {Peng},
  {Tonry}, {West}, \& {West}}]{2007ApJ...671.1456C}
{C{\^o}t{\'e}}, P. {et~al.} 2007, \apj, 671, 1456

\bibitem[{{C{\^o}t{\'e}} {et~al.}(2006){C{\^o}t{\'e}}, {Piatek}, {Ferrarese},
  {Jord{\'a}n}, {Merritt}, {Peng}, {Ha{\c s}egan}, {Blakeslee}, {Mei}, {West},
  {Milosavljevi{\'c}}, \& {Tonry}}]{2006ApJS..165...57C}
---. 2006, \apjs, 165, 57

\bibitem[{{Cox} {et~al.}(2006{\natexlab{a}}){Cox}, {Dutta}, {Di Matteo},
  {Hernquist}, {Hopkins}, {Robertson}, \& {Springel}}]{2006ApJ...650..791C}
{Cox}, T.~J., {Dutta}, S.~N., {Di Matteo}, T., {Hernquist}, L., {Hopkins},
  P.~F., {Robertson}, B., \& {Springel}, V. 2006{\natexlab{a}}, \apj, 650, 791

\bibitem[{{Cox} {et~al.}(2006{\natexlab{b}}){Cox}, {Jonsson}, {Primack}, \&
  {Somerville}}]{2006MNRAS.373.1013C}
{Cox}, T.~J., {Jonsson}, P., {Primack}, J.~R., \& {Somerville}, R.~S.
  2006{\natexlab{b}}, \mnras, 373, 1013

\bibitem[{{Dasyra} {et~al.}(2006{\natexlab{a}}){Dasyra}, {Tacconi}, {Davies},
  {Genzel}, {Lutz}, {Naab}, {Burkert}, {Veilleux}, \&
  {Sanders}}]{2006ApJ...638..745D}
{Dasyra}, K.~M. {et~al.} 2006{\natexlab{a}}, \apj, 638, 745

\bibitem[{{Dasyra} {et~al.}(2006{\natexlab{b}}){Dasyra}, {Tacconi}, {Davies},
  {Naab}, {Genzel}, {Lutz}, {Sturm}, {Baker}, {Veilleux}, {Sanders}, \&
  {Burkert}}]{2006ApJ...651..835D}
---. 2006{\natexlab{b}}, \apj, 651, 835

\bibitem[{{Davies} {et~al.}(2001){Davies}, {Kuntschner}, {Emsellem}, {Bacon},
  {Bureau}, {Carollo}, {Copin}, {Miller}, {Monnet}, {Peletier}, {Verolme}, \&
  {de Zeeuw}}]{2001ApJ...548L..33D}
{Davies}, R.~L. {et~al.} 2001, \apjl, 548, L33

\bibitem[{{de Jong}(1996)}]{1996A&A...313..377D}
{de Jong}, R.~S. 1996, \aap, 313, 377

\bibitem[{{Dehnen} \& {Gerhard}(1994)}]{1994MNRAS.268.1019D}
{Dehnen}, W., \& {Gerhard}, O.~E. 1994, \mnras, 268, 1019

\bibitem[{{Dekel} \& {Cox}(2006)}]{2006MNRAS.370.1445D}
{Dekel}, A., \& {Cox}, T.~J. 2006, \mnras, 370, 1445

\bibitem[{{Di Matteo} {et~al.}(2005){Di Matteo}, {Springel}, \&
  {Hernquist}}]{2005Natur.433..604D}
{Di Matteo}, T., {Springel}, V., \& {Hernquist}, L. 2005, \nat, 433, 604

\bibitem[{{di Serego Alighieri} {et~al.}(2006){di Serego Alighieri}, {Lanzoni},
  \& {J{\o}rgensen}}]{2006ApJ...647L..99D}
{di Serego Alighieri}, S., {Lanzoni}, B., \& {J{\o}rgensen}, I. 2006, \apjl,
  647, L99

\bibitem[{{Dressler}(1980)}]{1980ApJ...236..351D}
{Dressler}, A. 1980, \apj, 236, 351

\bibitem[{{Efstathiou}(2000)}]{2000MNRAS.317..697E}
{Efstathiou}, G. 2000, \mnras, 317, 697

\bibitem[{{Eggen} {et~al.}(1962){Eggen}, {Lynden-Bell}, \&
  {Sandage}}]{1962ApJ...136..748E}
{Eggen}, O.~J., {Lynden-Bell}, D., \& {Sandage}, A.~R. 1962, \apj, 136, 748

\bibitem[{{Emsellem} {et~al.}(2004){Emsellem}, {Cappellari}, {Peletier},
  {McDermid}, {Bacon}, {Bureau}, {Copin}, {Davies}, {Krajnovi{\'c}},
  {Kuntschner}, {Miller}, \& {de Zeeuw}}]{2004MNRAS.352..721E}
{Emsellem}, E. {et~al.} 2004, \mnras, 352, 721

\bibitem[{{Faber} \& {Jackson}(1976)}]{1976ApJ...204..668F}
{Faber}, S.~M., \& {Jackson}, R.~E. 1976, \apj, 204, 668

\bibitem[{{Ferrarese} {et~al.}(2006){Ferrarese}, {C{\^o}t{\'e}}, {Jord{\'a}n},
  {Peng}, {Blakeslee}, {Piatek}, {Mei}, {Merritt}, {Milosavljevi{\'c}},
  {Tonry}, \& {West}}]{2006ApJS..164..334F}
{Ferrarese}, L. {et~al.} 2006, \apjs, 164, 334

\bibitem[{{Fukugita} {et~al.}(1998){Fukugita}, {Hogan}, \&
  {Peebles}}]{1998ApJ...503..518F}
{Fukugita}, M., {Hogan}, C.~J., \& {Peebles}, P.~J.~E. 1998, \apj, 503, 518

\bibitem[{{Genzel} {et~al.}(1998){Genzel}, {Lutz}, {Sturm}, {Egami}, {Kunze},
  {Moorwood}, {Rigopoulou}, {Spoon}, {Sternberg}, {Tacconi-Garman}, {Tacconi},
  \& {Thatte}}]{1998ApJ...498..579G}
{Genzel}, R. {et~al.} 1998, \apj, 498, 579

\bibitem[{{Genzel} {et~al.}(2001){Genzel}, {Tacconi}, {Rigopoulou}, {Lutz}, \&
  {Tecza}}]{2001ApJ...563..527G}
{Genzel}, R., {Tacconi}, L.~J., {Rigopoulou}, D., {Lutz}, D., \& {Tecza}, M.
  2001, \apj, 563, 527

\bibitem[{{Gerhard}(1981)}]{1981MNRAS.197..179G}
{Gerhard}, O.~E. 1981, \mnras, 197, 179

\bibitem[{{Gerhard} \& {Binney}(1985)}]{1985MNRAS.216..467G}
{Gerhard}, O.~E., \& {Binney}, J. 1985, \mnras, 216, 467

\bibitem[{{Gonz{\' a}lez-Garc{\'{\i}}a} \&
  {Balcells}(2005)}]{2005MNRAS.357..753G}
{Gonz{\' a}lez-Garc{\'{\i}}a}, A.~C., \& {Balcells}, M. 2005, \mnras, 357, 753

\bibitem[{{Gonz{\'a}lez-Garc{\'{\i}}a}
  {et~al.}(2006){Gonz{\'a}lez-Garc{\'{\i}}a}, {Balcells}, \&
  {Olshevsky}}]{2006MNRAS.372L..78G}
{Gonz{\'a}lez-Garc{\'{\i}}a}, A.~C., {Balcells}, M., \& {Olshevsky}, V.~S.
  2006, \mnras, 372, L78

\bibitem[{{Gonz{\'a}lez-Garc{\'{\i}}a} \& {van
  Albada}(2005)}]{2005MNRAS.361.1043G}
{Gonz{\'a}lez-Garc{\'{\i}}a}, A.~C., \& {van Albada}, T.~S. 2005, \mnras, 361,
  1043

\bibitem[{{Heavens} {et~al.}(2004){Heavens}, {Panter}, {Jimenez}, \&
  {Dunlop}}]{2004Natur.428..625H}
{Heavens}, A., {Panter}, B., {Jimenez}, R., \& {Dunlop}, J. 2004, \nat, 428,
  625

\bibitem[{{Hernquist}(1987)}]{1987ApJS...64..715H}
{Hernquist}, L. 1987, \apjs, 64, 715

\bibitem[{{Hernquist}(1989)}]{1989Natur.340..687H}
---. 1989, \nat, 340, 687

\bibitem[{{Hernquist}(1992)}]{1992ApJ...400..460H}
---. 1992, \apj, 400, 460

\bibitem[{{Hernquist} \& {Barnes}(1991)}]{1991Natur.354..210H}
{Hernquist}, L., \& {Barnes}, J.~E. 1991, \nat, 354, 210

\bibitem[{{Hernquist} \& {Spergel}(1992)}]{1992ApJ...399L.117H}
{Hernquist}, L., \& {Spergel}, D.~N. 1992, \apjl, 399, L117

\bibitem[{{Hernquist} {et~al.}(1993){Hernquist}, {Spergel}, \&
  {Heyl}}]{1993ApJ...416..415H}
{Hernquist}, L., {Spergel}, D.~N., \& {Heyl}, J.~S. 1993, \apj, 416, 415

\bibitem[{{Heyl} {et~al.}(1994){Heyl}, {Hernquist}, \&
  {Spergel}}]{1994ApJ...427..165H}
{Heyl}, J.~S., {Hernquist}, L., \& {Spergel}, D.~N. 1994, \apj, 427, 165

\bibitem[{{Hogg} {et~al.}(2002){Hogg}, {Blanton}, {Strateva}, {Bahcall},
  {Brinkmann}, {Csabai}, {Doi}, {Fukugita}, {Hennessy}, {Ivezi{\' c}}, {Knapp},
  {Lamb}, {Lupton}, {Munn}, {Nichol}, {Schlegel}, {Schneider}, \&
  {York}}]{2002AJ....124..646H}
{Hogg}, D.~W. {et~al.} 2002, \aj, 124, 646

\bibitem[{{Hogg} {et~al.}(2003){Hogg}, {Blanton}, {Eisenstein}, {Gunn},
  {Schlegel}, {Zehavi}, {Bahcall}, {Brinkmann}, {Csabai}, {Schneider},
  {Weinberg}, \& {York}}]{2003ApJ...585L...5H}
---. 2003, \apjl, 585, L5

\bibitem[{{Hopkins} {et~al.}(2005){Hopkins}, {Hernquist}, {Cox}, {Di Matteo},
  {Martini}, {Robertson}, \& {Springel}}]{2005ApJ...630..705H}
{Hopkins}, P.~F., {Hernquist}, L., {Cox}, T.~J., {Di Matteo}, T., {Martini},
  P., {Robertson}, B., \& {Springel}, V. 2005, \apj, 630, 705

\bibitem[{{Hopkins} {et~al.}(2006){Hopkins}, {Hernquist}, {Cox}, {Di Matteo},
  {Robertson}, \& {Springel}}]{2006ApJS..163....1H}
{Hopkins}, P.~F., {Hernquist}, L., {Cox}, T.~J., {Di Matteo}, T., {Robertson},
  B., \& {Springel}, V. 2006, \apjs, 163, 1

\bibitem[{{Hopkins} {et~al.}(2008){Hopkins}, {Hernquist}, {Cox}, {Dutta}, \&
  {Rothberg}}]{2008arXiv0802.0508H}
{Hopkins}, P.~F., {Hernquist}, L., {Cox}, T.~J., {Dutta}, S.~N., \& {Rothberg},
  B. 2008, ArXiv e-prints, 802

\bibitem[{{Hubble}(1937)}]{1937C&T....53..194H}
{Hubble}, E. 1937, Ciel et Terre, 53, 194

\bibitem[{{Hubble} \& {Humason}(1931)}]{1931ApJ....74...43H}
{Hubble}, E., \& {Humason}, M.~L. 1931, \apj, 74, 43

\bibitem[{{Jesseit} {et~al.}(2005){Jesseit}, {Naab}, \&
  {Burkert}}]{2005MNRAS.360.1185J}
{Jesseit}, R., {Naab}, T., \& {Burkert}, A. 2005, \mnras, 360, 1185

\bibitem[{{Jesseit} {et~al.}(2006){Jesseit}, {Naab}, {Peletier}, \&
  {Burkert}}]{2006astro.ph..6144J}
{Jesseit}, R., {Naab}, T., {Peletier}, R., \& {Burkert}, A. 2006,
  astro-ph/0606144

\bibitem[{{Johansson} \& {Efstathiou}(2006)}]{2006MNRAS.371.1519J}
{Johansson}, P.~H., \& {Efstathiou}, G. 2006, \mnras, 371, 1519

\bibitem[{{Johansson} {et~al.}(2008){Johansson}, {Naab}, \&
  {Burkert}}]{2008arXiv0802.0210J}
{Johansson}, P.~H., {Naab}, T., \& {Burkert}, A. 2008, ArXiv e-prints, 802

\bibitem[{{Kennicutt}(1998)}]{1998ApJ...498..541K}
{Kennicutt}, R.~C. 1998, \apj, 498, 541

\bibitem[{{Khochfar} \& {Burkert}(2003)}]{2003ApJ...597L.117K}
{Khochfar}, S., \& {Burkert}, A. 2003, \apjl, 597, L117

\bibitem[{{Khochfar} \& {Burkert}(2005)}]{2005MNRAS.359.1379K}
---. 2005, \mnras, 359, 1379

\bibitem[{{Khochfar} \& {Silk}(2006{\natexlab{a}})}]{2006ApJ...648L..21K}
{Khochfar}, S., \& {Silk}, J. 2006{\natexlab{a}}, \apjl, 648, L21

\bibitem[{{Khochfar} \& {Silk}(2006{\natexlab{b}})}]{2006MNRAS.370..902K}
---. 2006{\natexlab{b}}, \mnras, 370, 902

\bibitem[{{Kormendy}(1989)}]{1989ApJ...342L..63K}
{Kormendy}, J. 1989, \apjl, 342, L63

\bibitem[{{Kormendy} \& {Bender}(1996)}]{1996ApJ...464L.119K}
{Kormendy}, J., \& {Bender}, R. 1996, \apjl, 464, L119+

\bibitem[{{Kormendy} \& {Djorgovski}(1989)}]{1989ARA&A..27..235K}
{Kormendy}, J., \& {Djorgovski}, S. 1989, \araa, 27, 235

\bibitem[{{Kriek} {et~al.}(2006){Kriek}, {van Dokkum}, {Franx}, {Quadri},
  {Gawiser}, {Herrera}, {Illingworth}, {Labb{\'e}}, {Lira}, {Marchesini},
  {Rix}, {Rudnick}, {Taylor}, {Toft}, {Urry}, \& {Wuyts}}]{2006ApJ...649L..71K}
{Kriek}, M. {et~al.} 2006, \apjl, 649, L71

\bibitem[{{Larson}(1969)}]{1969MNRAS.145..405L}
{Larson}, R.~B. 1969, \mnras, 145, 405

\bibitem[{{Larson}(1974)}]{1974MNRAS.169..229L}
---. 1974, \mnras, 169, 229

\bibitem[{{Larson}(1975)}]{1975MNRAS.173..671L}
---. 1975, \mnras, 173, 671

\bibitem[{{Li} {et~al.}(2006){Li}, {Hernquist}, {Robertson}, {Cox}, {Hopkins},
  {Springel}, {Gao}, {Di Matteo}, {Zentner}, {Jenkins}, \&
  {Yoshida}}]{2006astro.ph..8190L}
{Li}, Y. {et~al.} 2006, ArXiv Astrophysics e-prints

\bibitem[{{Mao} {et~al.}(1998){Mao}, {Mo}, \& {White}}]{1998MNRAS.297L..71M}
{Mao}, S., {Mo}, H.~J., \& {White}, S.~D.~M. 1998, \mnras, 297, L71

\bibitem[{{McDermid} {et~al.}(2006){McDermid}, {Emsellem}, {Shapiro}, {Bacon},
  {Bureau}, {Cappellari}, {Davies}, {de Zeeuw}, {Falc{\'o}n-Barroso},
  {Krajnovi{\'c}}, {Kuntschner}, {Peletier}, \& {Sarzi}}]{2006MNRAS.373..906M}
{McDermid}, R.~M. {et~al.} 2006, \mnras, 373, 906

\bibitem[{{McGaugh} \& {de Blok}(1997)}]{1997ApJ...481..689M}
{McGaugh}, S.~S., \& {de Blok}, W.~J.~G. 1997, \apj, 481, 689

\bibitem[{{McIntosh} {et~al.}(2005){McIntosh}, {Zabludoff}, {Rix}, \&
  {Caldwell}}]{2005ApJ...619..193M}
{McIntosh}, D.~H., {Zabludoff}, A.~I., {Rix}, H.-W., \& {Caldwell}, N. 2005,
  \apj, 619, 193

\bibitem[{{Melnick} \& {Sargent}(1977)}]{1977ApJ...215..401M}
{Melnick}, J., \& {Sargent}, W.~L.~W. 1977, \apj, 215, 401

\bibitem[{{Merritt} \& {Fridman}(1996)}]{1996ApJ...460..136M}
{Merritt}, D., \& {Fridman}, T. 1996, \apj, 460, 136

\bibitem[{{Mihos} \& {Hernquist}(1996)}]{1996ApJ...464..641M}
{Mihos}, J.~C., \& {Hernquist}, L. 1996, \apj, 464, 641

\bibitem[{{Naab} \& {Burkert}(2001)}]{2001ApJ...555L..91N}
{Naab}, T., \& {Burkert}, A. 2001, \apjl, 555, L91

\bibitem[{{Naab} \& {Burkert}(2003)}]{2003ApJ...597..893N}
---. 2003, \apj, 597, 893

\bibitem[{{Naab} {et~al.}(2006{\natexlab{a}}){Naab}, {Jesseit}, \&
  {Burkert}}]{2006MNRAS.372..839N}
{Naab}, T., {Jesseit}, R., \& {Burkert}, A. 2006{\natexlab{a}}, \mnras, 372,
  839

\bibitem[{{Naab} {et~al.}(2007){Naab}, {Johansson}, {Ostriker}, \&
  {Efstathiou}}]{2007ApJ...658..710N}
{Naab}, T., {Johansson}, P.~H., {Ostriker}, J.~P., \& {Efstathiou}, G. 2007,
  \apj, 658, 710

\bibitem[{{Naab} {et~al.}(2006{\natexlab{b}}){Naab}, {Khochfar}, \&
  {Burkert}}]{2006ApJ...636L..81N}
{Naab}, T., {Khochfar}, S., \& {Burkert}, A. 2006{\natexlab{b}}, \apjl, 636,
  L81

\bibitem[{{Naab} \& {Ostriker}(2006)}]{2006MNRAS.366..899N}
{Naab}, T., \& {Ostriker}, J.~P. 2006, \mnras, 366, 899

\bibitem[{{Naab} \& {Trujillo}(2006)}]{2006MNRAS.369..625N}
{Naab}, T., \& {Trujillo}, I. 2006, \mnras, 369, 625

\bibitem[{{Nagamine} {et~al.}(2006){Nagamine}, {Ostriker}, {Fukugita}, \&
  {Cen}}]{2006ApJ...653..881N}
{Nagamine}, K., {Ostriker}, J.~P., {Fukugita}, M., \& {Cen}, R. 2006, \apj,
  653, 881

\bibitem[{{Nakamura} {et~al.}(2003){Nakamura}, {Fukugita}, {Yasuda}, {Loveday},
  {Brinkmann}, {Schneider}, {Shimasaku}, \& {SubbaRao}}]{2003AJ....125.1682N}
{Nakamura}, O., {Fukugita}, M., {Yasuda}, N., {Loveday}, J., {Brinkmann}, J.,
  {Schneider}, D.~P., {Shimasaku}, K., \& {SubbaRao}, M. 2003, \aj, 125, 1682

\bibitem[{{Negroponte} \& {White}(1983)}]{1983MNRAS.205.1009N}
{Negroponte}, J., \& {White}, S.~D.~M. 1983, \mnras, 205, 1009

\bibitem[{{Ostriker}(1980)}]{1980ComAp...8..177O}
{Ostriker}, J.~P. 1980, Comments on Astrophysics, 8, 177

\bibitem[{{Ostriker} \& {Tinsley}(1975)}]{1975ApJ...201L..51O}
{Ostriker}, J.~P., \& {Tinsley}, B.~M. 1975, \apjl, 201, L51+

\bibitem[{{Partridge} \& {Peebles}(1967)}]{1967ApJ...147..868P}
{Partridge}, R.~B., \& {Peebles}, P.~J.~E. 1967, \apj, 147, 868

\bibitem[{{Pipino} \& {Matteucci}(2006)}]{2006MNRAS.365.1114P}
{Pipino}, A., \& {Matteucci}, F. 2006, \mnras, 365, 1114

\bibitem[{{Postman} \& {Geller}(1984)}]{1984ApJ...281...95P}
{Postman}, M., \& {Geller}, M.~J. 1984, \apj, 281, 95

\bibitem[{{Prantzos} \& {Silk}(1998)}]{1998ApJ...507..229P}
{Prantzos}, N., \& {Silk}, J. 1998, \apj, 507, 229

\bibitem[{{Rix} {et~al.}(1999){Rix}, {Carollo}, \&
  {Freeman}}]{1999ApJ...513L..25R}
{Rix}, H., {Carollo}, C.~M., \& {Freeman}, K. 1999, \apjl, 513, L25

\bibitem[{{Rix} \& {White}(1990)}]{1990ApJ...362...52R}
{Rix}, H.-W., \& {White}, S.~D.~M. 1990, \apj, 362, 52

\bibitem[{{Robertson} {et~al.}(2006{\natexlab{a}}){Robertson}, {Cox},
  {Hernquist}, {Franx}, {Hopkins}, {Martini}, \&
  {Springel}}]{2006ApJ...641...21R}
{Robertson}, B., {Cox}, T.~J., {Hernquist}, L., {Franx}, M., {Hopkins}, P.~F.,
  {Martini}, P., \& {Springel}, V. 2006{\natexlab{a}}, \apj, 641, 21

\bibitem[{{Robertson} {et~al.}(2006{\natexlab{b}}){Robertson}, {Hernquist},
  {Cox}, {Di Matteo}, {Hopkins}, {Martini}, \&
  {Springel}}]{2006ApJ...641...90R}
{Robertson}, B., {Hernquist}, L., {Cox}, T.~J., {Di Matteo}, T., {Hopkins},
  P.~F., {Martini}, P., \& {Springel}, V. 2006{\natexlab{b}}, \apj, 641, 90

\bibitem[{{Robin} {et~al.}(2003){Robin}, {Reyl{\' e}}, {Derri{\` e}re}, \&
  {Picaud}}]{2003A&A...409..523R}
{Robin}, A.~C., {Reyl{\' e}}, C., {Derri{\` e}re}, S., \& {Picaud}, S. 2003,
  \aap, 409, 523

\bibitem[{{Rothberg} \& {Joseph}(2004)}]{2004AJ....128.2098R}
{Rothberg}, B., \& {Joseph}, R.~D. 2004, \aj, 128, 2098

\bibitem[{{Rothberg} \& {Joseph}(2006)}]{2006astro.ph..4493R}
---. 2006, arXiv:astro-ph/0604493

\bibitem[{{Salpeter}(1955)}]{1955ApJ...121..161S}
{Salpeter}, E.~E. 1955, \apj, 121, 161

\bibitem[{{Schwarzschild}(1993)}]{1993ApJ...409..563S}
{Schwarzschild}, M. 1993, \apj, 409, 563

\bibitem[{{Scorza} {et~al.}(1998){Scorza}, {Bender}, {Winkelmann},
  {Capaccioli}, \& {Macchetto}}]{1998A&AS..131..265S}
{Scorza}, C., {Bender}, R., {Winkelmann}, C., {Capaccioli}, M., \& {Macchetto},
  D.~F. 1998, \aaps, 131, 265

\bibitem[{{Searle} {et~al.}(1973){Searle}, {Sargent}, \&
  {Bagnuolo}}]{1973ApJ...179..427S}
{Searle}, L., {Sargent}, W.~L.~W., \& {Bagnuolo}, W.~G. 1973, \apj, 179, 427

\bibitem[{{Springel} {et~al.}(2005){Springel}, {Di Matteo}, \&
  {Hernquist}}]{2005ApJ...620L..79S}
{Springel}, V., {Di Matteo}, T., \& {Hernquist}, L. 2005, \apjl, 620, L79

\bibitem[{{Surma} \& {Bender}(1995)}]{1995A&A...298..405S}
{Surma}, P., \& {Bender}, R. 1995, \aap, 298, 405

\bibitem[{{Thomas} {et~al.}(2005){Thomas}, {Maraston}, {Bender}, \& {de
  Oliveira}}]{2005ApJ...621..673T}
{Thomas}, D., {Maraston}, C., {Bender}, R., \& {de Oliveira}, C.~M. 2005, \apj,
  621, 673

\bibitem[{{Toomre}(1974)}]{1974IAUS...58..347T}
{Toomre}, A. 1974, in IAU Symp. 58: The Formation and Dynamics of Galaxies, 347

\bibitem[{{Toomre}(1977)}]{1977egsp.conf..401T}
{Toomre}, A. 1977, in Evolution of Galaxies and Stellar Populations, 401

\bibitem[{{Toomre} \& {Toomre}(1972)}]{1972ApJ...178..623T}
{Toomre}, A., \& {Toomre}, J. 1972, \apj, 178, 623

\bibitem[{{Treu} {et~al.}(2005){Treu}, {Ellis}, {Liao}, {van Dokkum}, {Tozzi},
  {Coil}, {Newman}, {Cooper}, \& {Davis}}]{2005ApJ...633..174T}
{Treu}, T. {et~al.} 2005, \apj, 633, 174

\bibitem[{{Trujillo} {et~al.}(2006){Trujillo}, {F{\"o}rster Schreiber},
  {Rudnick}, {Barden}, {Franx}, {Rix}, {Caldwell}, {McIntosh}, {Toft},
  {H{\"a}ussler}, {Zirm}, {van Dokkum}, \& {Labb{\'e}}}]{2006ApJ...650...18T}
{Trujillo}, I. {et~al.} 2006, \apj, 650, 18

\bibitem[{{Trujillo} \& {Pohlen}(2005)}]{2005ApJ...630L..17T}
{Trujillo}, I., \& {Pohlen}, M. 2005, \apjl, 630, L17

\bibitem[{{Valluri} \& {Merritt}(1998)}]{1998ApJ...506..686V}
{Valluri}, M., \& {Merritt}, D. 1998, \apj, 506, 686

\bibitem[{{van der Wel} {et~al.}(2005){van der Wel}, {Franx}, {van Dokkum},
  {Rix}, {Illingworth}, \& {Rosati}}]{2005ApJ...631..145V}
{van der Wel}, A., {Franx}, M., {van Dokkum}, P.~G., {Rix}, H.-W.,
  {Illingworth}, G.~D., \& {Rosati}, P. 2005, \apj, 631, 145

\bibitem[{{van Dokkum} \& {Franx}(1996)}]{1996MNRAS.281..985V}
{van Dokkum}, P.~G., \& {Franx}, M. 1996, \mnras, 281, 985

\bibitem[{{van Dokkum} {et~al.}(1998){van Dokkum}, {Franx}, {Kelson}, \&
  {Illingworth}}]{1998ApJ...504L..17V}
{van Dokkum}, P.~G., {Franx}, M., {Kelson}, D.~D., \& {Illingworth}, G.~D.
  1998, \apjl, 504, L17+

\bibitem[{{van Dokkum} {et~al.}(2008){van Dokkum}, {Franx}, {Kriek}, {Holden},
  {Illingworth}, {Magee}, {Bouwens}, {Marchesini}, {Quadri}, {Rudnick},
  {Taylor}, \& {Toft}}]{2008ApJ...677L...5V}
{van Dokkum}, P.~G. {et~al.} 2008, \apjl, 677, L5

\bibitem[{{van Dokkum} {et~al.}(2006){van Dokkum}, {Quadri}, {Marchesini},
  {Rudnick}, {Franx}, {Gawiser}, {Herrera}, {Wuyts}, {Lira}, {Labb{\'e}},
  {Maza}, {Illingworth}, {F{\"o}rster Schreiber}, \& {et
  al.}}]{2006ApJ...638L..59V}
---. 2006, \apjl, 638, L59

\bibitem[{{White} \& {Rees}(1978)}]{1978MNRAS.183..341W}
{White}, S.~D.~M., \& {Rees}, M.~J. 1978, \mnras, 183, 341

\bibitem[{{Whitmore} \& {Gilmore}(1991)}]{1991ApJ...367...64W}
{Whitmore}, B.~C., \& {Gilmore}, D.~M. 1991, \apj, 367, 64

\end{thebibliography}

\end{document}